\documentclass{aa}
\input{psfig}
\usepackage{graphicx}
\usepackage{amsmath}
\usepackage{amssymb}

\begin{document}

\title{Building Giant-Planet Cores at a Planet Trap}

\author{A. Morbidelli\inst{1}, A. Crida\inst{2}, F. Masset\inst{3,}\inst{4},
  R.P. Nelson\inst{5}}

\titlerunning{Building planetary cores at a planet trap}
\authorrunning{A. Morbidelli \it et el.}
\offprints{A. Morbidelli}

\institute{Observatoire de la C\^ote d'Azur, B.P. 4229, 06304 Nice
           Cedex 4, FRANCE\\ \email{morby@obs-nice.fr} \and Institut
           f\"ur Astronomie \& Astrophysik, Univ. of Tubingen, Germany
           \and Laboratoire AIM, CEA/Saclay, France \and 
IA-UNAM, Mexico City \and Queen Mary, University of
           London, England}

\abstract{A well-known bottleneck for the core-accretion model of
  giant-planet formation is the loss of the cores into the star by
  Type-I migration, due to the tidal interactions with the gas disk.
  It has been shown that a steep surface-density gradient in the disk,
  such as the one expected at the boundary between an active and a
  dead zone, acts as a planet trap and prevents isolated cores from
  migrating down to the central star.}  {We study the relevance of the
  planet trap concept for the accretion and evolution of systems of
  multiple planetary embryos/cores.}  {We performed hydrodynamical
  simulations of the evolution of systems of multiple massive objects
  in the vicinity of a planet trap. The planetary embryos evolve in 3
  dimensions, whereas the disk is modeled with a 2D grid. Synthetic
  forces are applied onto the embryos to mimic the damping effect that
  the disk has on their inclinations.}  {Systems with two embryos tend
  to acquire stable, separated and non-migrating orbits, with the more
  massive embryo placed at the planet trap and the lighter one farther
  out in the disk. Systems of multiple embryos are intrinsically
  unstable. Consequently, a long phase of mutual scattering can lead
  to accreting collisions among embryos; some embryos are injected
  into the inner part of the disk, where they can be evacuated into
  the star by Type~I migration. The system can resume a stable,
  non-migrating configuration only when the number of surviving
  embryos decreases to a small value ($\sim$~2-4). This can explain
  the limited number of giant planets in our solar system. These
  results should apply in general to any case in which the Type-I
  migration of the inner embryo is prevented by some mechanism, and
  not solely to the planet trap scenario.}

\maketitle

\section{Introduction}

The core-accretion model for the formation of giant planets meets a
serious problem. The cores of the planets, embedded in the gas disk,
should undergo Type~I migration towards the central star (Ward, 1997;
Tanaka et al., 2002). The migration rate increases linearly with the
core's mass. For masses of order of a few Earth masses, the
timescale for the engulfment of the core into the star is much shorter
than that for the accretion of a massive atmosphere and the transition
to the status of a giant planet (Pollack et al., 1996). Even the most
recent models of giant-planet formation, accounting for the positive
feedback of migration on accretion (Alibert et al., 2005), require
Type~I migration rates that are at least 10 times slower than
estimated by analytic theory and measured in hydrodynamical
simulations for cores embedded in laminar disks.

The situation is very different in turbulent disks (Nelson and
Papaloizou, 2003; 2004; Nelson, 2005; Johnson et al., 2006). The large scale 
fluctuations of the disk surface density under MRI turbulence
exert a stochastic torque onto the core. Consequently, the evolution
of the core's semi major axis resembles a random walk. For at least
some of the cores, the dynamical lifetime against collisions with the
star can increase substantially relative to the non-turbulent case,
dominated by Type~I migration (Nelson, 2005). 
However, the problem is that the stochastic torque exerted by turbulent
fluctuations also excites the orbital eccentricities of cores and
planetesimals. This inhibits accretion because mutual planetesimal
collisions become disruptive and runaway growth (which is
effective only when the core's escape velocity is large relative to
the velocity dispersion of the planetesimal population) is triggered
only for bodies more massive than about Ceres or Pluto. Thus, we are
confronted with a conundrum. If we invoke turbulence to solve the
core's Type~I migration problem, we cannot grow the core in first
place; if we invoke the core's growth in a low-turbulent portion of
the disk (the so-called `dead zone'; Gammie, 1996; Stone et al., 1998), then we
cannot avoid Type~I migration. 

Some possibilities have been proposed in order to avoid Type~I
migration in non-turbulent (e.g. laminar) disks.  Masset et
al. (2006b) found that, due to non-linear effects, the migration rate
is strongly reduced (or even halted) for a core of about 10-20 Earth
masses ($M_\oplus$), the exact value depending on disk
parameters. Paardekooper and Mellema (2006) found that inward type I
 migration could be reduced or reversed in non-isothermal discs,
 due to changes in the gas density in regions leading and
 trailing the planet. 

The `planet trap' concept, proposed by Masset et al. (2006a), provides
an alternative, appealing possibility, that we explore more in detail in this
paper. The planet trap results from the balance between the (negative)
differential Lindblad torque, usually responsible for Type~I migration
(Ward, 1997), and the (positive) coorbital corotation torque that is
exerted if the planet is in a portion of the disk characterized by a
steep and positive density gradient in the radial direction.
The name `planet trap' comes from the fact that 
planetary embryos,  migrating towards the central star, 
stop when they encounter the aforementioned density gradient, as if they were
captured  into a trap (see figures in Masset et al., 2006a or
Fig.~\ref{init} in this work). Notice that a positive density
gradient in the disk also acts as a trap for small planetesimals
suffering gas-drag. In fact, elsewhere in the disk the gas orbits
around the star with sub-keplerian speed because of a negative pressure
gradient. This forces the planetesimals to drift inwards. However, 
at a positive density gradient, the pressure gradient is also
positive, so that the gas is super-Keplerian. Consequently,  
planetesimals migrating inward because of gas drag would get
trapped at the point where the rotation profile transitions
between sub and super-Keplerian values. This would clearly help
the buildup of a large core located at/near the planet trap.

A density gradient responsible for the onset of a planet trap 
is expected to exist at the transition between
the inner, turbulent portion of the disk and the dead zone, because
the dead zone is characterized by a smaller transport coefficient (or
turbulent viscosity). This opens the intriguing possibility that giant-planet cores form in the dead zone (where the dispersion velocities
are low and accretion can be effective) but are not lost by Type~I
migration because of the planet trap at the inner edge of the dead
zone. 

The location of the inner edge of the dead zone is very
uncertain. Models of disk chemistry (see for instance Ilgner and
Nelson, 2006a) suggest that only the region interior to ~0.2 AU is
active (due to thermal ionization of potassium), with the dead zone
lying outside this region. However, models accounting for the
diffusion of free charges from the upper, ionized layers of the disk
down to the mid-plane (Ilgner and Nelson, 2006b; Turner et al., 2007)
suggest that the magneto-rotational instability fills the whole
thickness of the disk up to at least 5 AU, under favourable
circumstances.  Moreover, Chiang and Murray-Clay (2007) argued that
the boundary between the active zone and the dead zone moves outward
with time.  A gas density gradient suitable for the creation of a
planet trap could also be caused by the partial depletion of the inner
few AUs of the disk by winds driven by the magnetic field (Ferreira et
al., 2006). Notice that there is a growing body of evidence for the
existence of transition disks with inner cavities of several AUs in
size (see for instance Forrest et al, 2004; Calvet et al., 2004, 2005;
Bergin et al., 2004; Pietu et al., 2006).  So, the location of the
planet trap is an open issue, but it is not excluded that it can be
several AUs away from the central star. By considering that the planet
trap might have played a role in the formation of the cores of the
giant planets of our solar system, we implicitly assume that it was
located at about 5~AU.

The location and effectiveness of the planet trap depend on the disk
properties, but are the same for a large variety of embryo's
masses. Thus, all planetary embryos tend to go to the same location
(see for instance Fig. 3 by Masset et al., 2006a). Masset et
al. conjectured that, for this reason, the planet trap might be a
sweet spot for the rapid accretion of a massive planetary core from
mutual collisions of embryos.  However, they did not test this
conjecture with numerical simulations. Actually, Fig.~3 by Masset et
al. is very misleading.  The figure superposes the evolution of many
embryos of multiple masses, each of which integrated
separately. Neither direct nor indirect perturbations among the
embryos were taken into account. Thus, the clustering of embryos at
the planet trap can be artificial.

Terquem and Papaloizou (2007) took the opposite attitude concerning
the relevance of the planet trap. They argued that, in a multi-embryo
system, embryo-embryo scattering could move the semi major axis of an
embryo across the density gradient into the inner disk, where Type~I
migration would resume again. Moreover they speculated that the
interaction among the embryos at the vicinity of the trap would excite
their orbital eccentricities; embryos on eccentric orbits would only
sample the density gradient for a fraction of the orbit and,
consequently, they would feel a reduced corotation torque.

In summary, the actual role of the planet trap for the accretion and
preservation of giant-planet cores from a system of multiple
planetary embryos is not clear, and different outcomes can be
legitimately expected. The purpose of this paper is to conduct
specific hydrodynamical simulations of the dynamics of massive bodies
in the vicinity of the planet trap. In section~\ref{code} we start by
presenting the numerical scheme that we use in the simulations and in
section~\ref{setup} we describe our set-up for building the required
surface density gradient and identifying the location of the planet
trap. We then proceed in sequence, making the problem more complex step by
step. In section~\ref{2pl} we investigate the dynamics of two
planetary embryos, initially on distant orbits, as a function of their
masses and mass ratio. In section~\ref{scatt}, we study the dynamics
of two embryos on unstable orbits, scattering off each other. Finally,
in section~\ref{multiple}, we consider a system with many (i.e. 10)
planetary embryos, initially of equal masses. 

Several simulations of the dynamics of multiple proto-planets in a gas
disk have been presented in the literature.  Papaloizou and
Szuszkiewicz (2005) studied the the migration-induced resonance
trapping in a system of two planets with masses in the Earth mass
range. McNeil et al. (2005), Cresswell and Nelson (2006) and Terquem
and Papaloizou (2007) studied the migration of multiple
proto-planetary embryos. These studies used either N-body integrators
with fake forces that mimic the planet-disk interactions (McNeil et
al., 2005; Terquem and Papaloizou, 2007) or fully hydrodynamical
simulations (Papaloizou and Szuszkiewicz, 2005) or both (Cresswell and
Nelson, 2006). In all cases, a classical laminar disk was considered,
without a planet trap. The general result is that, after an initial
phase during which some mutual collisions may be possible, the
proto-planets find a relative configuration that is stable, each
object locked in resonance with another. The full system
collectively evolves by Type~I migration towards the central star. 
The difference between our results and this general evolution will
enlighten the role of the putative planet trap. 

\section{An hybrid numerical simulation scheme}  
\label{code}

Because some of our simulations will involve mutual scattering and
collisions among planetary embryos, it is important that the
simulations are done in three dimensions, namely allowing the orbits
to evolve also in inclination. In fact, it is well known that in a
co-planar system the ratio between collisions and scattering events is
artificially large (Chambers, 2001), which invalidates the first
generation of planet accretion simulations done in two dimensions.
However, doing a fully three-dimensional simulation of the evolution
of the embryos and the disk would be prohibitively expensive, from the
computational point of view. For this reason we have adopted a
compromise approach, where the embryos are allowed to evolve in
three dimensions, the disk is simulated with a two-dimensional grid,
and fake damping forces are applied on embryo inclinations. 

Our hybrid (2D+3D) implementation is built on the code FARGO by 
Masset (2000a, 2000b). Each massive body is identified by its three
dimensional positions ($x,y,z$) and velocities ($v_x,v_y,v_z$). 
As the disk is planar, each cell is identified by the position of its
center ($x_c,y_c$). The potential of the body-cell gravitational
interaction is therefore:
\begin{equation}
U=-{{GMm}\over{\sqrt{(x-x_c)^2+(y-y_c)^2+z^2+\epsilon^2}}}\ ,
\label{pot}
\end{equation}
where $G$ is the gravitational constant (assumed to be 1), $M$ is the
mass of the body (in solar masses), $m$ is the mass of the gas in the
cell and $\epsilon$ is the so-called smoothing parameter, here assumed
to be 70\% of the local thickness of the disk ($H/r=0.05$ in all
simulations presented in this paper). The smoothing parameter is
intended to mimic the thickness of the disk and not the inclination of
the planet. The functional form (\ref{pot}) has been chosen in order
to have an analytic function that tends to the usual potential adopted in
2D simulations for vanishing $z$, and to the correct Newtonian 
potential for $z>>\epsilon$.  
Moreover, because the smoothing parameter that we adopt is larger
than the Hill radius and the Bondi radius of the embryos, we account
for the torque exerted on the planet by every cell of disk (e.g. with
no torque cut-off; see Masset et al., 2006b for a discussion).

\begin{figure}[h!]
%\centerline{\psfig{figure=../pl_trap/trap2_8_6_4_2_6x1/plot_ei.ps,height=6.cm}
\centerline{\psfig{figure=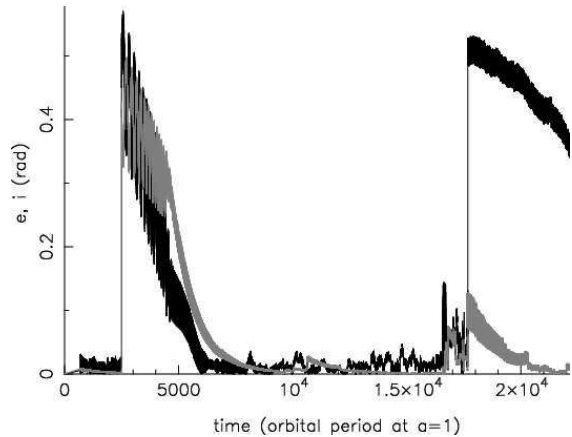,height=5.7cm}}
\vspace*{-.1cm}
\caption{The evolution of the eccentricity (black curve) and
  inclination (in radians; gray curve) of an embryo undergoing scattering from
  other objects in the system. Notice the good agreement between the
  damping timescales of eccentricity and inclination after each major
  scattering event.}
\label{damping}
\end{figure}

With this potential, the planetary bodies suffer migration and
eccentricity damping, but the inclination is not affected because the
disk cannot support vertical waves. In reality, the disk also 
damps the planet's orbital inclination (Lubow and Ogilvie,
2001). Thus, following Tanaka and Ward (2004), we mimic this damping
effect by exerting an acceleration on the planet's $v_z$ equal to
\begin{equation}
F_{{\rm damp},z}=M\left({{\Sigma_g}\over{c_s^4}}\right)\Omega(2A_z^c
v_z+ A_z^s z \Omega) \,
\label{i-damp}
\end{equation} 
where $\Sigma_g$ is the mean surface density of the gas on an annulus
located at the planet's current radial distance, $c_s$ is the local
sound speed and $\Omega$ is the keplerian angular velocity. The
constants $A_z^c$ and $A_z^s$ are given by Tanaka and Ward (2004).
Tanaka and Ward's theory is linear, and valid for small inclinations
(and eccentricities). More recently, Cresswell et al. (2007) studied
the damping of eccentricity and inclination with 3D numerical
simulations and found that, for large initial values of $e$ and/or
$i$, the damping is slower than predicted by the linear
theory. However, the similarity between the damping rates of $e$ and
$i$ still holds.  Thus, here we tune empirically the coefficients
$A_z^c$ and $A_z^s$ in (\ref{i-damp}) by a multiplicative factor, so
that the resulting damping on the inclination occurs in most occasions
on the same timescale of the damping of the eccentricity, which is
self-consistently computed by our hydrodynamical code. After several
tests, the multiplicative factor was set equal to 0.1 (i.e. Tanaka and
Ward formula was divided by a factor of 10). As an example,
Fig.~\ref{damping} shows the resulting evolution of the eccentricity
and of the inclination after two scattering events in one simulation
that will be described in sect.~\ref{multiple}. With this choice of
the multiplicative factor, the damping timescales of both quantities
essentially coincide. We acknowledge that, for small inclinations, the
real damping rate of the inclination should be faster than the one
that we impose. A faster inclination damping rate would increase
somewhat the probability that embryos undergo mutual collisions, in
particular in the cases where the scattering phase is protracted (as
in the multi-embryo simulation of section~\ref{multiple}). Because in
this paper we do not intend to study the planet accretion rate from a
quantitative point of view, but rather to investigate the qualitative
aspects of the dynamics of systems of embryos in the vicinity of a
planet trap, we think that having inclination damping rates that might
be smaller in some cases than the real ones should not be a severe
limitation.

As in FARGO, the gravitational interactions among the massive bodies
and their interactions with the star are integrated with a fixed
time-step Runge Kutta integrator (Steiner, 1996) of 5th order. The
choice of a fixed time-step is quite mandatory in the architecture of
the code, and, in principle, it may fail to give an accurate evolution
in case of very close encounters among the embryos. However, the
time-step, imposed by the so-called CFL condition (Masset, 2000a) for
the correct simulation of the gas dynamics, is very small, of order of
1/180 of the orbital period at our unit of distance ($a=1$), so that
the simulation should be correct in most cases. Again, we stress that
we are more interested in this work on the qualitative dynamics of embryos in
the vicinity of the planet trap, rather than in its quantitative
aspects (which would not be accurate anyway given the limitations of
the two dimensional treatment of the gas disk).

Our implementation of the Runge Kutta algorithm searches for physical
collisions among the embryos. This is done by computing, at each
time-step, the hyperbolic arc of an embryo relative to each other one,
and by confronting the minimal approach distance along this arc with
the physical sizes of the bodies. For the physical radii, we assume
for simplicity that the bulk density of the objects is 2g/cm$^3$ and
that our unit distance is the astronomical unit. If the unit of
distance or the bulk density were larger (smaller), 
the physical collisions would result less (more) frequent than in our
simulations.  We assume that all
collisions are accretional. When two embryos collide, one embryo is
suppressed from the simulation, while the other one acquires the sum
of the masses and its velocity is changed so that the linear momentum
of the two colliding embryos is carried by the surviving one.

\section{Simulation set up}
\label{setup}

In all our simulations, the disk extends from 0.4 to 3 in radius, and
the 2D grid has resolution 260x450 in radius and azimuth,
respectively. The disk aspect ratio is 5\% at all radii.  To set a
surface density profile with a positive gradient around $a=1$ we
proceed as follows (see also Masset et al., 2006a).  The viscosity of
the disk is set as a function of radius. More precisely, inside
$r=0.9$ the viscosity $\nu$ is given with an $\alpha$ prescription (Shakura
and Sunyaev, 1973), with $\alpha=6\times 10^{-3}$. Between $r=0.9$ and
$r=1.1$, the viscosity  decreases linearly as
$$
\nu=\nu(r=0.9)\times(4.857-r\times4.2857)\ .
$$ 
In the disk beyond $r=1.1$, the viscosity is given again with an
$\alpha$ prescription, but with $\alpha$ decreased by a factor
$4.857-1.1\times 4.2857=0.143$ with respect to the inner disk.

\begin{figure}[h!]
\vskip 10pt
\centerline{\psfig{figure=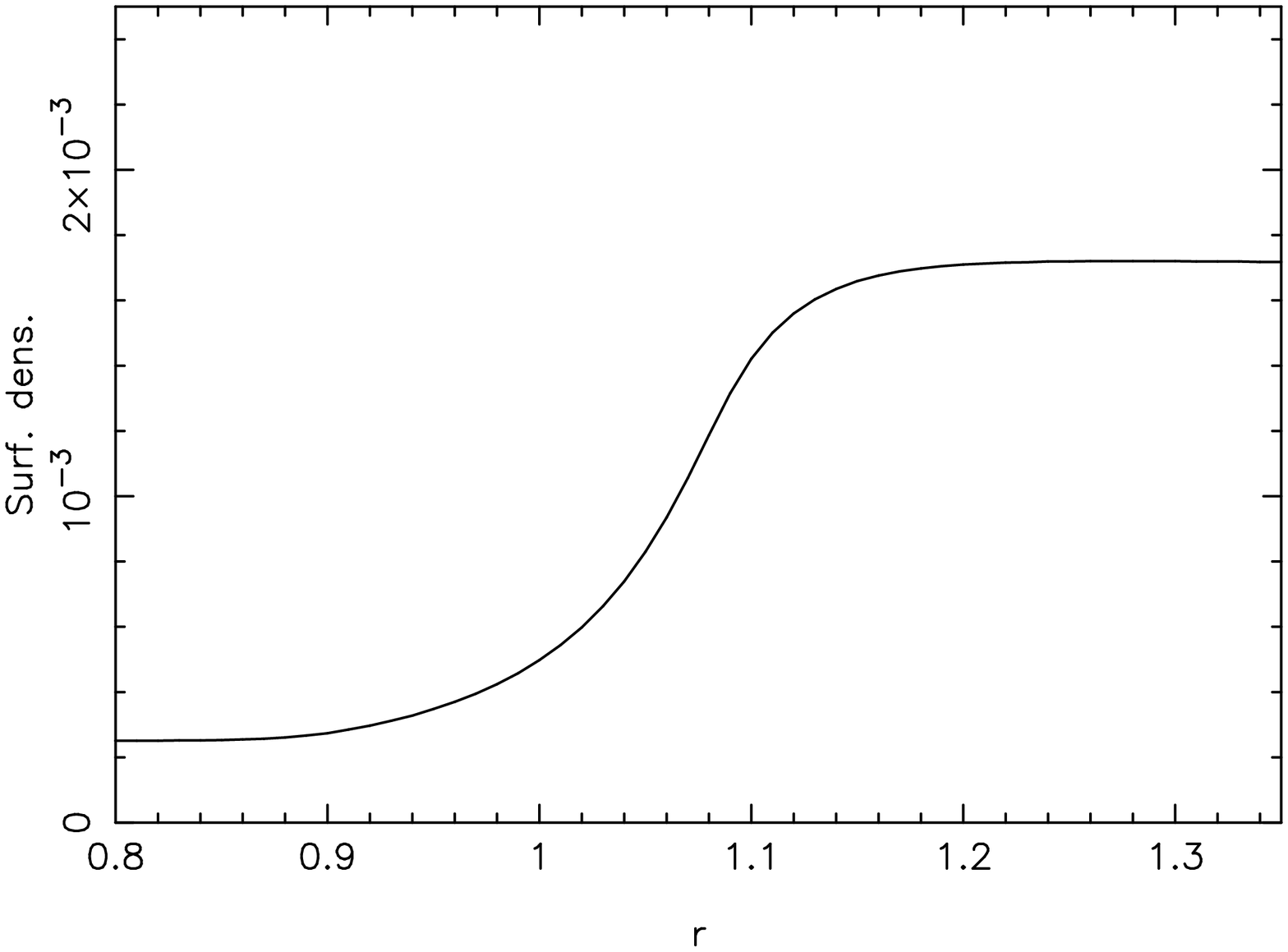,height=5.7cm}}
%\centerline{\psfig{figure=../pl_trap/trap2_10Me/plot_a.ps,height=6.cm}}
\vskip 15pt
\centerline{\psfig{figure=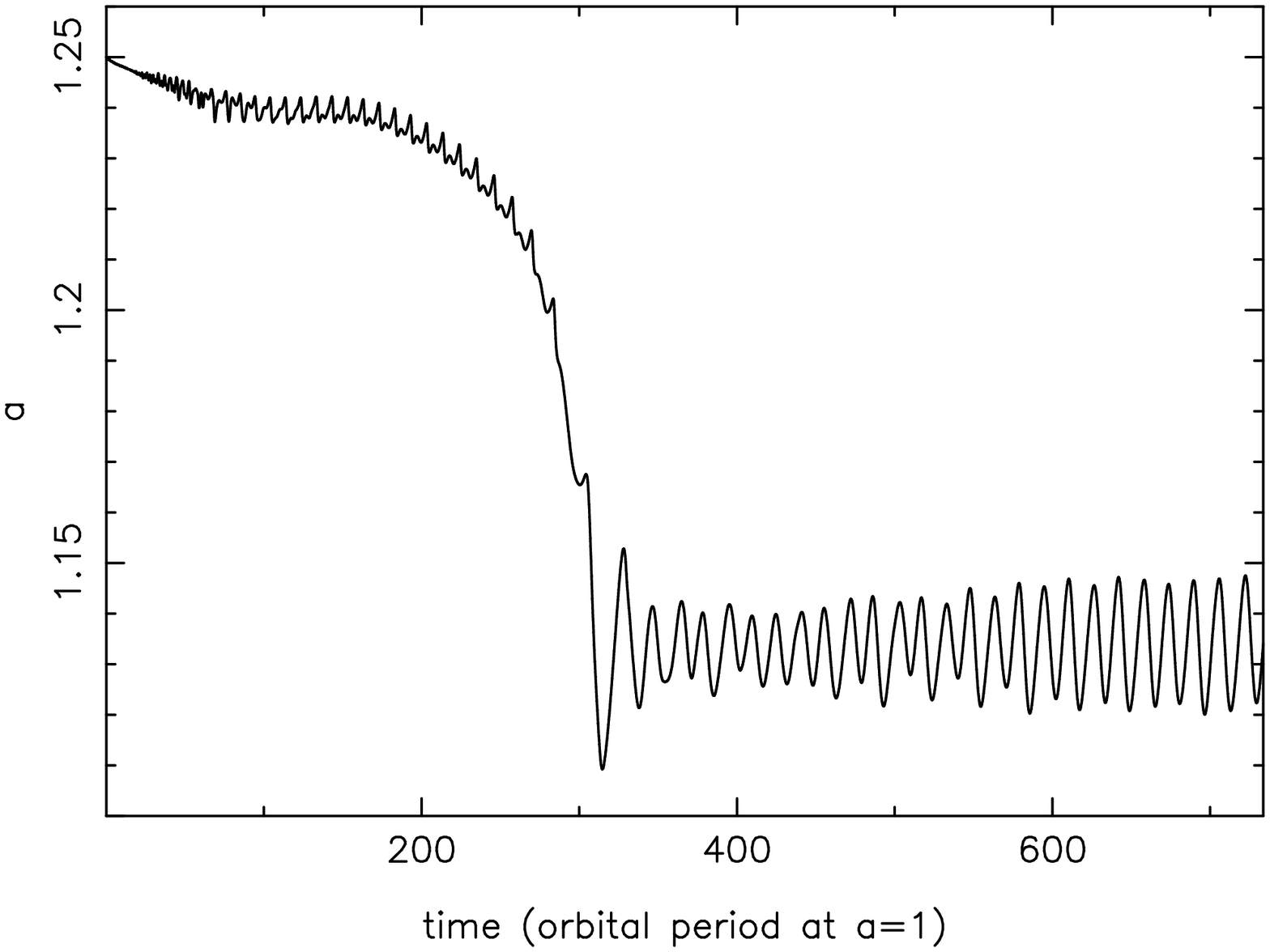,height=5.7cm}}
\vspace*{-.1cm}
\caption{Top: the equilibrium surface density distribution of the
  disk, with the steep surface density gradient at $r=1$, obtained
  with the viscosity and boundary conditions prescription illustrated in
  the text. Bottom: the evolution of the semi major axis of a
  10~$M_\oplus$ object, in the disk shown on the left panel. The
  object is rapidly captured into the planet trap. This simulation is
  used to determine the location of the trap, for the subsequent
  runs.}
\label{init}
\end{figure}

In a first simulation, we let the disk evolve for 15,000 orbits at
$a=1$, without any planetary body in it. The initial surface density
distribution increases linearly from $2.5\times 10^{-4}$ at $r=0.4$ to
$1.7\times 10^{-3}$ at $r=3$ in the units specified above.  The
boundary conditions preserve the surface density at the border of the
grid, and act as sources or sinks of mass. Under this setting and the
viscosity prescription described above, the disk evolves rapidly to a
new equilibrium configuration, illustrated in top panel of
Fig.~\ref{init}, that exhibits the desired surface density gradient.

We then use the final gas distribution of this simulation as an input
for a new simulation, in which we release a 10~$M_\oplus$ core at
$r=1.25$, with a small (e.g. $\sim 10^{-3}$) eccentricity and a null
inclination. The core evolves inward rapidly and gets caught at the
planet trap (see bottom panel of Fig.~\ref{init}). Notice the
acceleration of the inward migration, as the core approaches the trap,
already discussed by Masset et al. (2006a). Also, the evolution of the
semi major axis is not monotonic, but presents large oscillations,
that become very pronounced after the capture in the trap. This is due
to the existence of a vortex, produced by a Rossby instability at the
summit of the surface density gradient, where the vortensity has a
local maximum (Masset et al., 2006a). The planet trap is located
inside the orbital radius of the vortex, at a distance equal to a few
times $H/r$. Thus, the synodic period between the core and the vortex
is of order $P \times r/H$, where $P$ is the orbital period, and
corresponds to the oscillation period visible in Fig.~\ref{init}.

Because the planet trap mechanism is independent of the planet's mass,
in the simulations of the following sections, whenever we need to
place a planetary embryo in the trap, we simply adopt as initial
conditions the final output of the simulation in the bottom panel of
Fig.~\ref{init} and just reset the mass of the body.

\section{Dynamics of two embryos on initially separated orbits}
\label{2pl}

\begin{figure}[h!]
%\centerline{\psfig{figure=../pl_trap/plot_2Me_1Me.ps,height=6.cm}}
%\centerline{\psfig{figure=../pl_trap/plot_10Me_5Me.ps,height=6.cm}}
\centerline{\psfig{figure=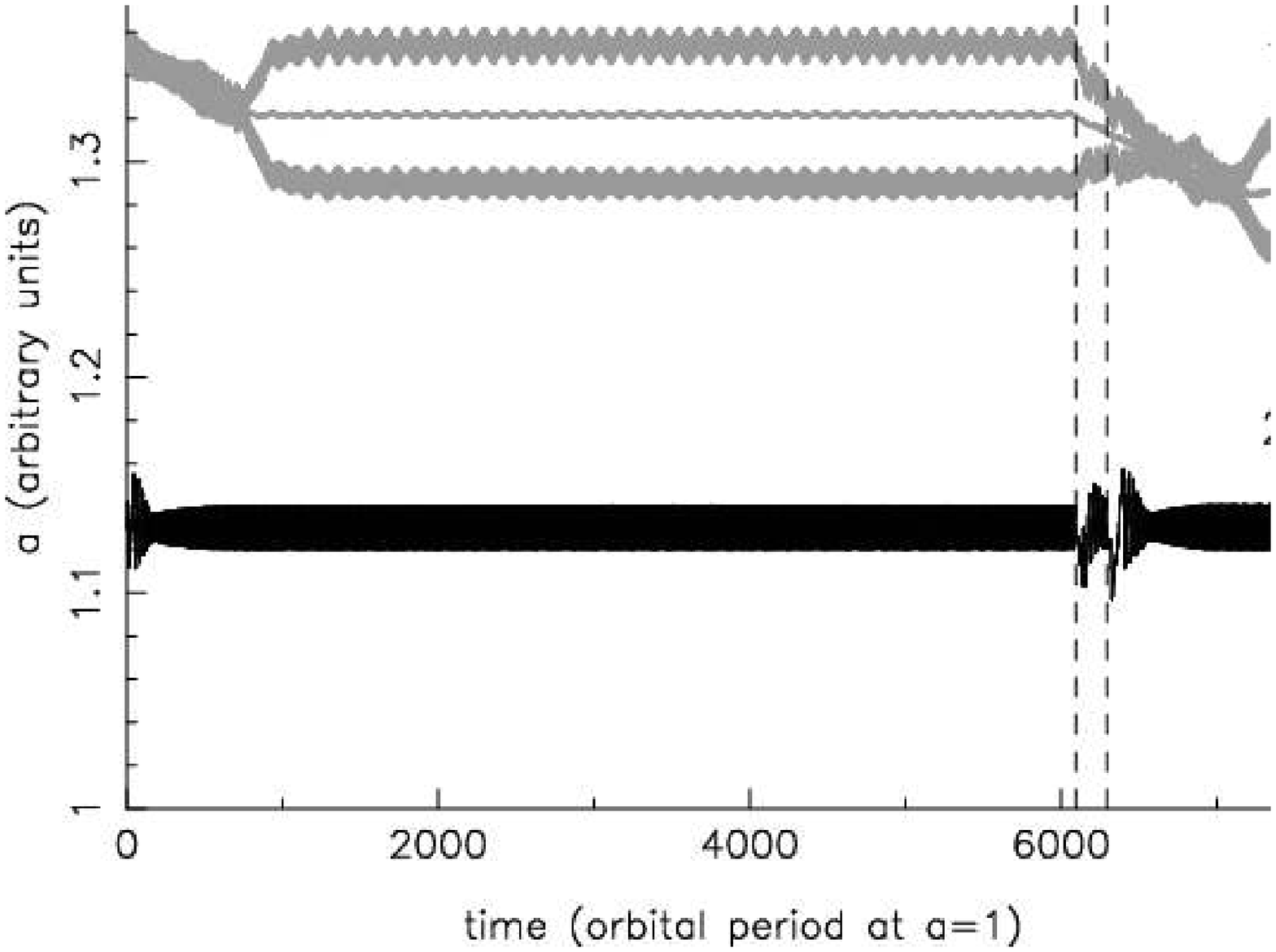,height=5.7cm}}
\vskip 15pt
\centerline{\psfig{figure=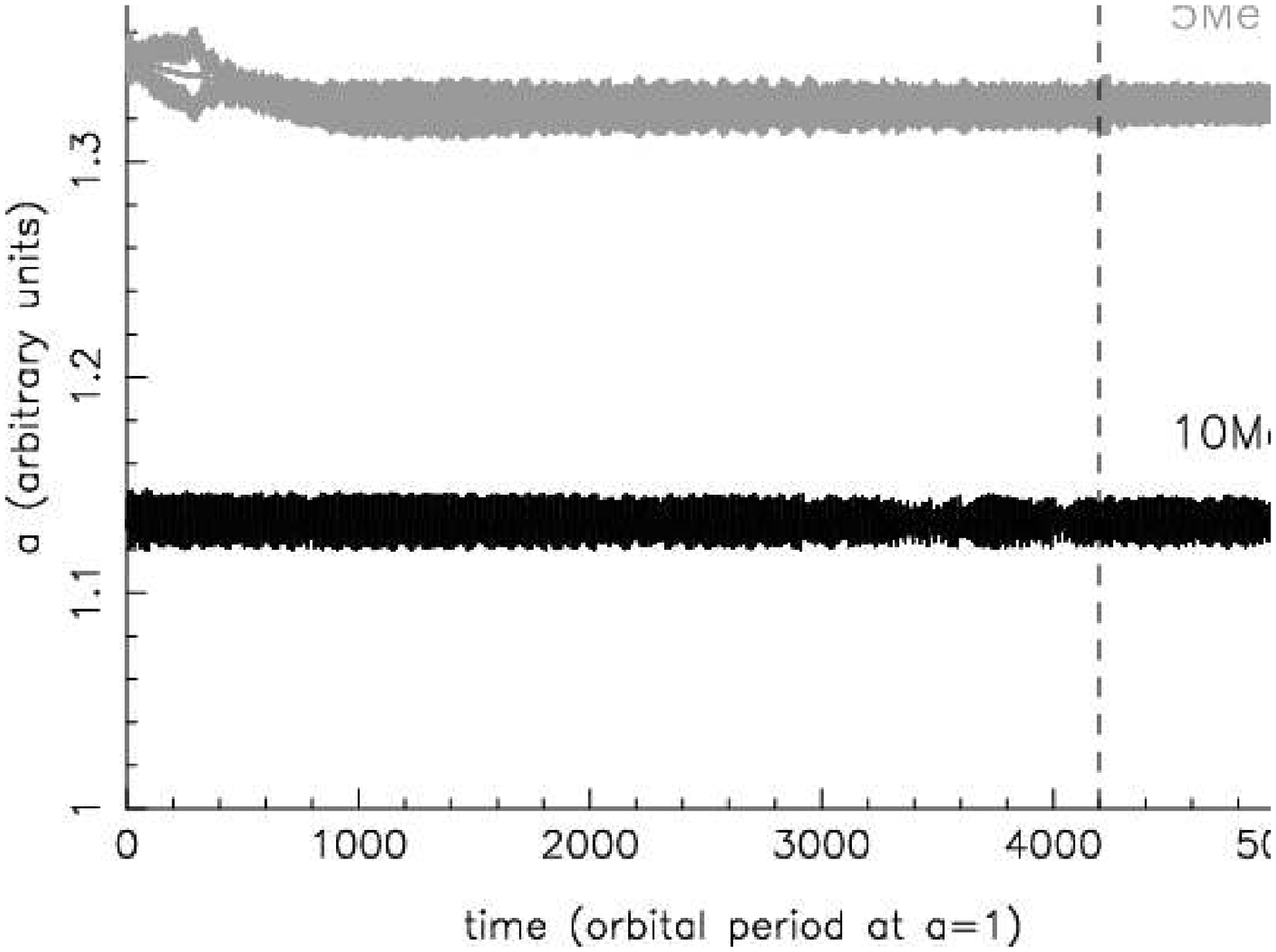,height=5.7cm}}
\vspace*{-.1cm}
\caption{The evolution of two embryos in presence of a planet
  trap. The outer body is represented by three curves, denoting perihelion
  distance, semi major axis and aphelion distance of its orbit. The
  inner embryo is initially placed at the trap. Because its eccentricity
  remains small only the evolution of its semi major axis is plotted. 
  Top panel: The masses
  of the inner and outer embryos are 2 and 1~$M_\oplus$
  respectively. Bottom panel: The masses of the inner and outer embryos
  are 10 and 5~$M_\oplus$ respectively. The vertical dashed lines show
  the times at which the direct perturbations between the embryos are
  switched off and on, respectively. See text for discussion.}
\label{dist}
\end{figure}

Our first set of simulations accounts for two planetary embryos, the
more massive of which is initially placed at the trap. The lighter
embryo is placed initially at $a=1.35$, on a quasi-circular orbit
($e\sim \times 10^{-3}$) with an initial inclination of $\sim 10^{-2}$
radians. We have done three simulations, with masses of the embryos
pairs equal to 10 and 5 $M_\oplus$, 5 and 2.5 $M_\oplus$, 2 and 1
$M_\oplus$.  Fig.~\ref{dist} shows two examples of
evolution, from the first and the last simulation. The outer object is
represented by three curves, denoting its perihelion distance, semi
major axis and aphelion distance, respectively, so that the evolution
of the eccentricity can also be captured by eye (when the orbit is
circular the three curves are superposed). For the inner embryo, only
the semi major axis is plotted versus time.

In all cases we observe that, after some inward migration, the outer
embryo stops. In none of the cases the outer embryo reaches the planet
trap. Thus, neither the conjecture by Masset et al. (2006a) -namely
that embryos cluster at the planet trap and accrete with each other-
nor that by Terquem and Papaloizou (2007) -that embryos scatter each
other off the trap, or become eccentric and insensitive to the planet
trap mechanism- appears to be true.

The mechanism that stops the inward migration of the outer embryo
depends on the embryo masses.  If the embryos are quite small as in
the 2--1~$M_\oplus$ simulation (as well as in the 5--2.5~$M_\oplus$
one, not illustrated here), the outer embryo is trapped in a mean
motion resonance with the inner one (the 4:5 resonance for the case
illustrated in the top panel of Fig.~\ref{dist}, for $1000<t<6000$). 
When resonance
trapping occurs, the eccentricity of the outer embryo is enhanced to a
non-zero equilibrium value. The direct perturbation of the inner
embryo onto the outer one is crucial in this dynamical evolution. To
illustrate this, after 6,050 orbits (time marked by the leftmost
vertical dashed line on the top panel of Fig.~\ref{dist}), we
suppress the direct perturbations between the embryos. Immediately,
the inward migration of the outer embryo starts again, and the
eccentricity is rapidly damped. After 6,200 orbits (time marked by the
rightmost vertical dashed line on the top panel of Fig.~\ref{dist}),
we switch on again the direct perturbations between the embryos. Then,
the outer embryo is captured into the next mean motion resonance (the
5:6 resonance), which leads to eccentricity excitation again.

Conversely, if the embryos are massive, as in the 10--5~$M_\oplus$
simulation, the outer embryo is repelled by the inner one through
indirect perturbations, that is through the modifications in the disk
surface density created by the first embryo.  To demonstrate the
dominant role of the indirect perturbation, again we suppress the
direct perturbations between the embryos after 4,200 orbital periods
(time marked by the vertical dashed line in the bottom panel of
Fig.~\ref{dist}). The evolution of the bodies does not change
significantly in this case. Notice also that the eccentricity of the
bodies remains small all the time.  Thus, the outer embryo is not
captured in a mean motion resonance with the inner one.

\begin{figure}[h!]
\centerline{\psfig{figure=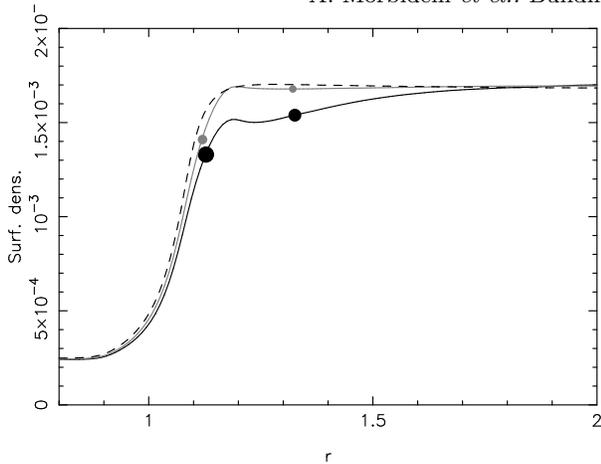,height=5.7cm}}
\vspace*{-.1cm}
\caption{Radial profiles of the azimuth-averaged disk surface
  densities in the case of the 10--5~$M_\oplus$ simulation (black
  curve), of the 2--1 ~$M_\oplus$ simulation (gray curve) and in
  absence of embryos (dashed curve). The
  radial location of each embryo is reported with a filled circle, with size
  proportional to the object's physical radius. The ordinate of each
  circle is arbitrary, and has been chosen so to please each embryo on
  the corresponding surface density curve, for illustrative purposes.}
\label{profiles}
\end{figure}

To investigate which kind of indirect perturbation repels the outer
embryo in this case, we compare in Fig.~\ref{profiles} the surface
density profile of the disk in this simulation (black curve) with the
one obtained in the 2--1~$M_\oplus$ simulation (gray curve).  In the
case of the 2--1~$M_\oplus$ simulation, the surface density profile is
essentially the unperturbed one, shown on the top panel of
Fig.~\ref{init}. The little `bump' visible at $r\sim 1.2$ is due to
the aforementioned vortex. The outer embryo is in a region of the disk
where the surface density profile is flat, so that it would suffer
Type~I migration unless it is restrained by a resonance with the inner
embryo.  In the case of the 10--5~$M_\oplus$ simulation, the surface
density profile is strongly modified.  The `bump' due to the vortex is
more pronounced.  Also, the torque exerted by the inner embryo on the
outer disk forces the surface density to acquire a positive gradient
at equilibrium in the 1.2--1.8 radial range. This produces a new
planet trap, in which the outer embryo is eventually captured.

In all the simulations described above, the inclination of the outer
embryo damps continuously and exponentially, down to values of order
$10^{-4}$ radians or smaller. This is due to the fact that close encounters
with the inner embryo do not occur, and that neither mean motion
resonance trapping nor the indirect perturbations excite the
inclination. Consequently, the damping effect exerted by the disk,
forced through the prescription (\ref{i-damp}), dominates the
evolution of the inclination. 

%The strong,
%indirect perturbation is probably due to the spiral wave launched in
%the disk by the inner embryo, which creates a density enhancement
%interior to the orbit of the outer embryo. This enhancement is
%localized in azimuth and on average, over a synodic period of the two
%embryos, can increase the inner Lindblad torque, without being
%compensated by the so-called `pressure buffer' (Ward, 1997).  This
%eventually sets an equilibrium between the inner and the outer
%Lindblad torques felt by the outer embryo, once the two embryos are
%close enough to each other.

The lesson that we derive from these experiments is that, at least in
these configurations, the planet trap is effective to prevent Type~I
migration of a system of two embryos. Close approaches and collisions
between embryos do not occur. Thus, the growth of the embryos can be
due only to the sole accretion of small planetesimal, as in a
classical oligarchic growth mode (Kokubo and Ida, 1998). The case of a
reversed mass ratio between the embryos (i.e. the case where the outer
one is the more massive) will be discussed in the second part of the
next section.

All the simulations illustrated above have been conducted in the
framework of a laminar disk. It is interesting to investigate how the
embryos' dynamics can be affected by large scale turbulent
fluctuations in the disk. Strong turbulence might in principle destroy
the planet trap mechanism, or inhibit resonant trapping and isolation
of the two massive bodies. Turbulence is expected to arise in the disk
due to Magneto-Rotational-Instabilities (MRI; e.g. Balbus and Hawley,
1991). Full MHD simulations are beyond the scope of this paper, but we
can mimic the effect of MRI turbulence adding a stochastic planar torque of
suitable amplitude on the embryos in our laminar hydrodynamical
simulations. The recipe that we follow for the generation of the
stochastic torque is that implemented by Ogihara et al. (2007; see
formul\ae\ 5-7 in that paper). In turn, Ogihara et al. implementation
is based on the scheme proposed in Laughlin et al. (2004), which was
calibrated on the results of MHD simulations. Notice, however, that
MRI turbulence does not only induce fluctuations in the gas density
distribution, but also in the gas velocity field. These velocity
fluctuations may affect the horseshoe streamlines in the corotation
region (Papaloizou et al. 2004), and therefore the
magnitude of the corotation torque.  Consideration of this effect goes
beyond the scope of this paper, so the fluctuations in the velocities
are neglected in these approximate implementations of the stochastic
torque.

\begin{figure}[h!]
%\centerline{\psfig{figure=../pl_trap/trap2_2Me_1Me_turb_gamma10-2/turb-gamma10-2.ps,height=6.cm}}
\vskip 10pt
\centerline{\psfig{figure=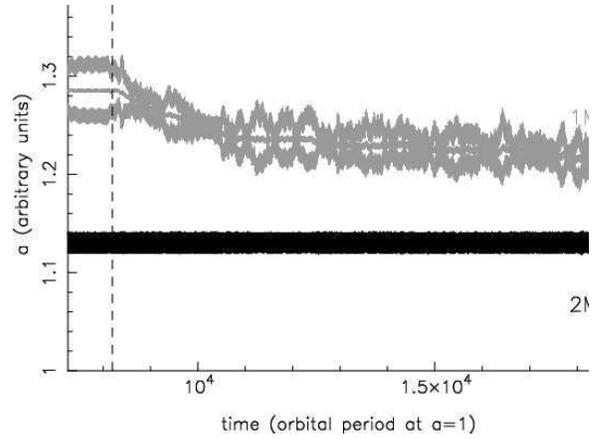,height=5.7cm}}
\vspace*{-.1cm}
\caption{The continuation of the simulation in the top panel of
  Fig.~\ref{dist}, but adding the effects of turbulent fluctuations in
  the gas surface density distribution of relative amplitude of about
  1\%. The vertical dashed line marks the time when the stochastic
  (turbulent) torque is switched on.}
\label{turb}
\end{figure} 

We have continued the simulation presented in the top panel of
Fig.~\ref{dist}, implementing the stochastic torque on {\it both}
embryos. We have done three simulations, with $\gamma=10^{-1}$,
$10^{-2}$ and $10^{-3}$, where $\gamma$ is the dimensionless parameter
identified by Ogihara et al. as indicator of the `turbulence
strength', corresponding approximately to the relative amplitude of
the fluctuations of the gas density field (Fig.~\ref{turb} presents
the result for $\gamma=10^{-2}$).  In all cases, the stochastic torque
does not destroy the planet trap mechanism. The inner embryo keeps
oscillating around $a=1.13$. This is in agreement with the similar
tests performed by Masset et al. (2006a), although with a simpler
recipe for the stochastic torque.  The trapping of the outer embryo in
the mean motion resonance is inhibited in the cases with
$\gamma=10^{-1}$ and $10^{-2}$. In fact, the embryo leaves the 5:6
resonance as soon as the stochastic torque is switched on, and is not
trapped permanently in any 
other mean motion resonance. Thus, it migrates stochastically towards
the inner embryo, until it has a close encounter with it. In the
example of Fig.~\ref{turb} the close encounter scatters the inner
embryo into the inner part of the disk, whereas the lighter embryos is
scattered onto an orbit with much larger semi major axis and
eccentricity (see however sect.~\ref{scatt} for a more detailed
discussion of the outcome of embryo-embryo scattering). In the case of
reduced turbulence strength ($\gamma=10^{-3}$), however, the outer
embryo is not released from the resonance. We do not show a figure for
this case, because the evolution of the system is indistinguishable
from that shown in the top panel of Fig.~\ref{dist}.

The MHD simulations of Laughlin et al. (2004) suggest that $\gamma$
should be of order $10^{-3}$ to $10^{-2}$ in the {\it active} zone of
the disk, although large uncertainties exist. In the dead zone,
however, the turbulence strength should be much smaller. Recall that
the planet trap should be positioned at the transition between the
active and the dead zone. Thus, the outer embryo, which is beyond the
planet trap location, should evolve in the dead zone.  Although waves
generated by the density fluctuation in the live zone can propagate
into the dead zone, we expect that the stochastic perturbations on the
outer embryo have a reduced intensity.  Therefore, from the
experiments above, we expect that it is unlikely that the outer embryo
could escape trapping in mean motion resonances with the inner
embryo. Notice also that the strength of mean motion resonances scales
with the square root of the embryo masses so that, for more massive
embryos, the isolation mechanism should be even more robust against
the effects of turbulence.

\section{Dynamics of two embryos on unstable, scattering orbits}
\label{scatt}

If the embryos are locked into a mutual mean motion resonance, their
orbital separation remains constant. However, if at the same time the
embryos are growing by accretion of planetesimals, they can
become too massive for their orbital separation to be stable. 
In fact, the minimal orbital separation that allows stability depends
on the masses of the bodies (Gladman, 1993). Thus -despite the
results of the previous section- we can expect that
the embryos eventually can achieve a phase of mutual scattering.

To explore the outcome of an instability phase in the vicinity of the
planet trap, we have done a new series of three simulations, in which
the inner embryo (2~$M_\oplus$) is placed at the trap, and the outer
one (1~$M_\oplus$) is placed at $a\sim 1.17$, close enough to the
inner embryo to be unstable from the beginning of the simulation. 
The initial eccentricity and inclination of the outer embryo are again
of order $10^{-3}$ and $10^{-2}$ radians, respectively. 

\begin{figure}[h!]
%\centerline{\psfig{figure=../pl_trap/plot_scatt_c_log.ps,height=6.cm}
\centerline{\psfig{figure=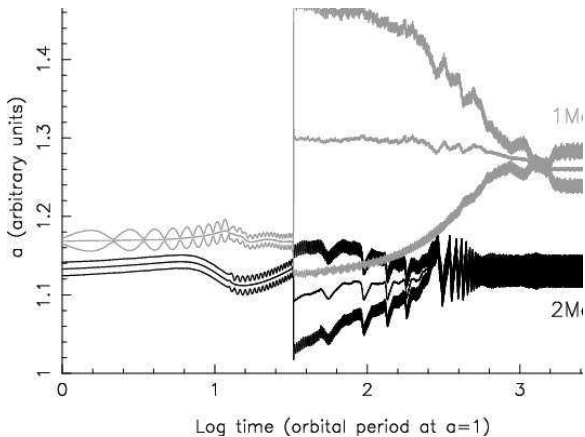,height=5.7cm}
}
\vspace*{-.1cm}
\caption{The evolution of two embryos on initially close, unstable
  orbits, in the vicinity of a planet trap.
  Each body is represented by three curves, denoting perihelion
  distance, semi major axis and aphelion distance of its orbit. 
  The time is plotted in logarithmic scale in order to enlighten the
  early phase of mutual scattering between the two embryos.}
\label{close}
\end{figure}

An example evolution from one of these simulations is illustrated in
Fig.~\ref{close}. At the first close encounter, the embryos acquire
eccentric orbits, that are much more separated in semi major axis than
the initial ones. Then, there is a competition between two processes:
the damping of the eccentricities of the embryos exerted by the disk
(Tanaka and Ward, 2004) and the excitation of the eccentricities due
to the repeated mutual encounters between the embryos. The first
process (eccentricity damping) is more effective.  Thus, the
eccentricities of both embryos decay, and the objects become
dynamically decoupled (no close encounters are possible any more). The
outer embryo starts to suffer an inward Type~I migration, until it is
trapped into a mean motion resonance (the 6:7 in this case) which, as
in the previous section, increases the eccentricity up to a new,
equilibrium value. A stable configuration is achieved, characterized
by an orbital separation larger than the one from which the simulation
was started (the final semi major axis of the outer embryo stabilizes
at 1.26, while it started at 1.17).

This kind of evolution is common to two out of three of our
simulations, and suggests that embryos can increase their orbital
separation as they grow, through short instability phases of mutual
scattering, followed by eccentricity damping and resonance trapping.
The situation is similar to that found in the simulations of runaway
growth of embryos in particle disks (Kokubo and Ida, 1996), where it
was also noticed that the mutual orbital separation of the embryos
increases with the embryos' masses. However, in one of our simulations
the evolution is different. The outer, lighter embryo is scattered
inward, into the inner part of the disk, as expected by Terquem and
Papaloizou (2007), where it will be free to migrate towards the
central star. In this case the planet trap retains only the more
massive embryo.
 
In all simulations, the scattering events excite the inclination of
 the more massive embryo to $\sim 0.05$ radians and the inclination of the
 less massive embryo to $\sim 0.1$ radians. After each scattering event, the
 inclination is damped exponentially, as imposed by eq.~(\ref{i-damp}).

\begin{figure}[h!]
%\centerline{\psfig{figure=../pl_trap/plot_1Me_2Me.ps,height=6.cm}
\centerline{\psfig{figure=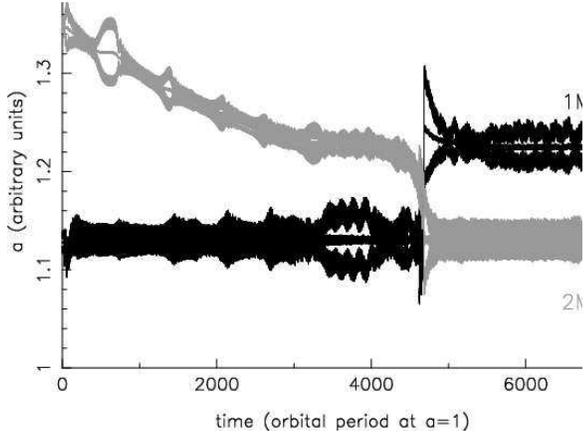,height=5.7cm}
}
\vspace*{-.1cm}
\caption{As in the left panel of Fig.~\ref{dist}, but for reversed
  values of the masses.}
\label{reversed}
\end{figure}

Because they involve some scattering, we present in this section also
simulations of two embryos initially on separated orbits as in
sect.~\ref{2pl}, but in which the more massive embryo is originally
the outer one. We have done three simulations of this kind, with
masses of the embryos pairs equal to 1 and 2 $M_\oplus$, 1 and 2.5
$M_\oplus$, 5 and 10 $M_\oplus$. An example evolution is given in
Fig.~\ref{reversed}, taken from the first of these simulations, but
common to all of them. Because it is less massive, the inner embryo
cannot prevent the migration of the outer embryo by trapping it in an
outer mean motion resonance. One could expect that the inner embryo is
pushed inwards from the trap by the outer embryo. However, this does
not occur, because the corotation torque at the planet trap is
stronger than the Lindblad torque felt by the outer embryo (see Fig.~1
in Masset et al., 2006a), despite the latter being more massive. What
happens is that the orbit of the inner embryo tends to become
eccentric when the outer one is trapped into a resonance. Eventually
the resonant configuration is broken, and the two embryos have close
encounters with each other.  A possible outcome of the close encounter
phase is a switch of position between the two bodies, with the more
massive embryo taking the position of the lighter one at the planet
trap. At this point, the evolution is analogous to that illustrated in
Fig.~\ref{close}. In some cases, the inner lighter embryo is scattered
into the inner disk, while the outer, more massive embryo is scattered
onto an orbit with large semi major axis and eccentricity. The
eccentricity of the orbit of the more massive embryo is eventually
damped and type I migration brings it into the planet trap. In none of
our three simulations we have observed a stable configuration with the
lighter embryo at the planet trap and the more massive embryo at
larger distance from the star.

\begin{figure}[h!]
%\centerline{\psfig{figure=../pl_trap/trap2_4cores_Tan/4cores.ps,height=6.cm}
\centerline{\psfig{figure=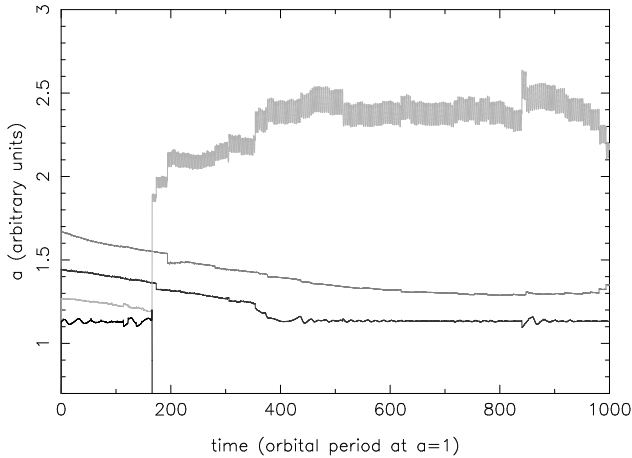,height=5.7cm}}
\vspace*{-.1cm}
\caption{The evolution of the semi major axes of 4 cores with masses
  equal to 5, 10, 14 and 14~$M_\oplus$ for the initially innermost to
  the outermost one. The 5~$M_\oplus$ core is initially placed at 
  the planet trap. The system is unstable and evolves to a
  configuration where one of the most massive cores takes the position
  at the trap.}
\label{4cores}
\end{figure}

Therefore, these results suggest that stable, non-migrating
configurations of embryos, in which the the innermost object is the
least massive one, should be rare. This is quite in contrast with the
mass hierarchy of the cores of the giant planets of our solar system,
where it seems (although the error bars are large) that the core of
Jupiter is less massive than that of Saturn, which is in turn less
massive than that of Uranus or Neptune (Guillot, 1999). As a
confirmation that a system of 4 cores with masses increasing with
heliocentric distance is unlikely to be stable, we placed a
5~$M_\oplus$ core at the planet trap ($a\sim 1.13$), one core of
10~$M_\oplus$ at $a\sim 1.27$ and two cores of 14~$M_\oplus$ at $a\sim
1.45$ and $a\sim 1.67$ respectively.  These core masses are consistent
with those inferred for the current cores of the giant planets, given
the large uncertainties.  We did 2 simulations, with slightly
different initial locations of the cores. Both showed similar
evolutions (see Fig.~\ref{4cores} for one example). The second embryo
pushed its way towards the planet trap, and eventually destabilized
the inner embryo.  The lightest embryo was scattered into the inner
part of the disk, and the second lightest one was scattered outward,
beyond the two most massive cores. One of the most massive cores
eventually took the place at the planet trap.

If the current cores of the giant planets have indeed masses that
increase with heliocentric distance, the experiments above suggest one
of the two possibilities: (i) either the core of Saturn completed its
growth after that Jupiter had already acquired a substantially massive
atmosphere, and similarly for Uranus relative to
Saturn or (ii) the core of Jupiter and (partially) that of Saturn have
been eroded by convective motion in the massive atmospheres of these
giant planets, and they were significantly more massive when they
formed (so that the mass hierarchy that allows stability was
respected).

\section{Dynamics of a system of multiple embryos}
\label{multiple}

The result of the previous sections show that a system with two
planetary embryos tends to find a stable configuration 
in which the bodies are isolated from each other. However, the
simulations with the 4 cores of the giant planets suggest that, with
increasing number of bodies, the evolution can be more
chaotic. Indeed, in more crowded systems of
planetary embryos, we may expect that the dynamics is very different, 
as embryos can have more difficulties in finding a mutual orbital spacing
that is large enough to ensure stability.

In this section we present a simulation with 10 embryos, each of one Earth
mass. The inner one is placed at the planet trap at $a\sim 1.13$. The
others are placed at increasing semi major axes, so that the initial
orbital separation between two adjacent embryos is equal to 5 mutual
Hill radii. The outermost planetary embryo is therefore placed at
$a=2$. The initial eccentricities of all embryos are of order
$10^{-3}$. The initial inclinations are of order $10^{-2}$ radians, and the 
longitudes of the ascending node are alternated by 180 degrees, relative
to the neighboring embryo.

\begin{figure}[h!]
%\centerline{\psfig{figure=../pl_trap/trap2_10_1Me_5Hill/plot.ps,height=6.cm}
\centerline{\psfig{figure=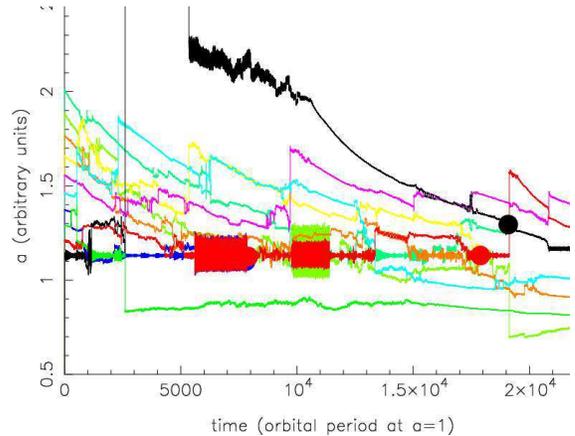,height=5.7cm}}
\vspace*{-.1cm}
\caption{The evolution of the semi major axes of 10 embryos of
  1~$M_\oplus$ each in the vicinity of the planet trap. The big red and
  black dots mark three collision events recorded in this simulation.}
\label{10pl}
\end{figure} 

The evolution of the embryos is illustrated in Fig.~\ref{10pl}.  As
expected, the system evolves much more chaotically than in the
previous simulations.  While the embryos at, or close to, the trap
tend not to migrate, the outermost ones drift towards the star by
Type~I migration. Some resonance trapping occurs, but these resonant
phases do not remain stable for long time. Mutual scattering therefore
dominates the dynamics and lasts for a much longer time than in the
simulations with only two embryos presented in the previous
sections. The embryo at the planet trap can be kicked out of the trap
and its place can be taken by another embryo. There are episodes where
two embryos share the same orbital semi major axis, in a relative
tadpole motion (for instance the brown and light blue embryos between
$t\sim$1,000 and $t\sim$2,000, or the brown and the red embryos
between $t\sim$13,500 and $t\sim$15,000). Planets on mutual tadpole
orbits have already been observed in the simulations by Cresswell and
Nelson (2006). We also find cases of embryos temporarily on a mutual
satellite motion (the red and the blue embryos between $t\sim$5,500
and $t\sim$7,500, and the red and green embryos between $t\sim$10,000
and $t\sim$11,500). These episodes are well visible, as they are
characterized by large oscillations of the semi major axes of the
concerned embryos. To our knowledge, this is the first simulation in
which planets are found to trap each other in satellite motion. A
satellite capture requires some kind of energy dissipation, presumably
due to the interaction of the objects with the disk. Given the limited
resolution of the grid used for the hydrodynamical calculation, we
cannot exclude that the dissipation is artificially large, so that
episodes of satellite motion might be much more rare in reality. 

Due to the protracted phase of mutual encounters, three mutual
collisions happen in this simulation. Two involve the embryo whose
evolution is shown in red, and one the embryo whose evolution is shown
in black. The collision events are marked by the big red and black
dots respectively. 

The evolution of the embryos calms down at $t\sim$21,000. At this
time, 4 of the 7 surviving embryos have been kicked into the inner
part of the disk ($a\le 1$). These embryos do not migrate
significantly, because the surface density of the inner disk in our
simulation is very low. If the inner disk had been more massive, these
embryos would have started a monotonic inward Type~I migration, as in
Cresswell and Nelson (2006) simulations, and would have eventually
been eliminated by a collision with the star. The black embryo (which
has grown to 2~$M_\oplus$) has taken the place at the planet trap. It
seems to prevent the migration of the red embryo (3~$M_\oplus$) and
magenta embryo (1~$M_\oplus$) outside of its orbit, similarly to what
we have seen in the previous sections. Because the red embryo is the
most massive, it might be possible that a continuation of this
simulation could show an exchange between the orbits of the red and
the black embryos, similarly to what happened to the red and magenta
embryos at $t\sim$21,000.

Throughout the evolution, the inclinations of the embryos are excited
by the scattering events, up to a few tenths of a radian, and then damp
exponentially, until the next scattering event occurs. Fig.~\ref{damping}
shows the evolution of the inclination of one of the embryos.

The comparison of this simulation with those of Cresswell and Nelson
(2006) is quite instructive. Here, because of the presence of the
planet trap, which is an effective obstacle to migration, the embryos
cannot find as easily a mutual stable orbital spacing. Thus the
scattering phase is much more protracted in time, and collisions are
more likely to occur. In addition, in Cresswell and Nelson
simulations, all the embryos eventually drift towards the central
star, while here three embryos are saved from migration, at the trap or
just beyond it. Therefore, the planet trap appears to be an effective
mechanism to prevent Type~I migration of some embryos, even in
originally crowded systems. In these systems, the presence of the trap
helps mutual collisions to occur and favors the growth of larger objects,
even if not in the way envisioned in Masset et al. (2006a). Also, the
trap acts like a filter, in the sense that, out of many initial
embryos, only a few are allowed to survive in a stable non-migrating
configuration. As long as the system is too crowded, stability cannot
be achieved, and the number of embryos has to decay through
collisions, ejections or injections into the inner disk. Eventually, a
single planet trap helps the formation and the preservation of a few
cores. This result may provide an explanation of why our
solar system has only a limited number of giant planets, and no
intermediate mass (of few Earth masses) planets in between them.

\section{Conclusions}

We have conducted a series of hydrodynamical simulations of the
evolution of planetary embryos in a gas disk. The simulation scheme
allows the embryos to evolve in three dimensions, whereas the disk is
modeled with a 2 dimensional grid. A damping force on the inclinations
of the embryos is applied. The disk has been built with a
positive density gradient at $r\sim 1$,  so that a `planet trap'
(Masset et al., 2006a) is produced at $r\sim 1.1$.

Our simulations show that, because the planet trap is an obstacle to
Type~I migration, a crowded system of embryos can not drift in concert
towards the central star, preserving the relative mutual
separations among the embryos. Instead, the system becomes violently
unstable. Collisions between embryos become possible. Some embryos can
be scattered into the inner part of the disk, where the planet trap
mechanism is not effective any more to prevent their migration.  The
system stabilizes when the number of embryos behind the planet trap
decreases down to a few. This may explain why our solar system has
only a small number of giant planets.

A system with two (probably a few) embryos tends to find a stable,
non-migrating configuration. The embryo at the planet trap restrains the
other embryo(s) from migrating, through the action of its outer mean
motion resonances, or (if its mass is large enough) through indirect
perturbations. In this configuration, the embryos can continue to
increase their mass in solid elements only by accreting small
planetesimals, in a classical oligarchic growth regime. If they become
too massive with respect to their orbital separation, they can acquire
new, more separated orbits that are again stable and non
migrating. This happens through
a short phase of instability, during which mutual encounters 
emplace the objects onto orbits with more separate semi major axes and
larger eccentricities, followed by aphase of eccentricity
damping exerted by the disk. During the instability phases, the more
massive embryo typically manages to take position at the planet trap,
the lighter embryo(s) being emplaced on an orbit at larger semi major
axis. Thus, the fact that in our solar system Jupiter is the giant-planet with the smallest core suggests that (i) either the completion
of the cores of Saturn, Uranus and Neptune occurred after Jupiter had
already acquired a substantial mass of gas in its envelope (possibly
opening a gap and moving the planet trap to the outer edge of its
gap), or (ii) that the the core of Jupiter (and partially that of
Saturn) was originally more massive and was substantially eroded by
convective motion in its atmosphere.

These results and considerations should not be limited to the planet trap
scenario. Their validity should be more general. They should apply to
any case in which some mechanism prevents an inner embryo to have a
free type-I migration.

\acknowledgements
{We thank Alice Quillen and Willy Kley for useful, friendly
  comments. A.M. is grateful for support from the French programs PNP and OPV.} 
\endacknowledgements

\section{References}

%\begin{itemize}

%\end{itemize}


\begin{thebibliography}{[AHU]}
\bibitem[]{} Alibert, Y., Mordasini, 
C., Benz, W., \& Winisdoerffer, C.\ 2005, \aap, 434, 343 
\bibitem[]{} Balbus, S.~A., \& 
Hawley, J.~F.\ 1991, \apj, 376, 214 
\bibitem[Bergin et al.(2004)]{2004ApJ...614L.133B} Bergin, E., et al.\ 
2004, \apjl, 614, L133 
\bibitem[Calvet et al.(2004)]{2004ASPC..324..205C} Calvet, N., Hartmann, 
L., Wilner, D., Walsh, A., \& Sitko, M.~L.\ 2004, Debris Disks and the 
Formation of Planets, 324, 205 
\bibitem[Calvet et al.(2005)]{2005ApJ...630L.185C} Calvet, N., et al.\ 
2005, \apjl, 630, L185 
\bibitem[]{} Chambers, J.~E.\ 2001, 
Icarus, 149, 262 
\bibitem[]{} Chiang, E.~I., 
\& Murray-Clay, R.~A.\ 2007, ArXiv e-prints, 706, arXiv:0706.1241 
\bibitem[]{} Cresswell, P., \& 
Nelson, R.~P.\ 2006, \aap, 450, 833 
\bibitem[]{} Cresswell, P., 
Dirksen, G., Kley, W., \& Nelson, R.~P.\ 2007, ArXiv e-prints, 707, 
arXiv:0707.2225 
\bibitem[]{} Ferreira, J., 
Dougados, C., \& Cabrit, S.\ 2006, \aap, 453, 785 
\bibitem[Forrest et al.(2004)]{2004ApJS..154..443F} Forrest, W.~J., et al.\ 
2004, \apjs, 154, 443 
\bibitem[]{} Gammie, C., 1996, ApJ, 457, 355 
\bibitem[]{} Gladman, B.\ 1993, Icarus, 
106, 247 
\bibitem[]{} Guillot, T.\ 1999, Science, 
286, 72 
\bibitem[]{} Ilgner, M., \& 
Nelson, R.~P.\ 2006a, \aap, 445, 223 
\bibitem[]{} Ilgner, M. Nelson, R.P., 2006b, A \& A, 445, 223 
\bibitem[]{} Johnson, E.~T., 
Goodman, J., \& Menou, K.\ 2006, \apj, 647, 1413 
\bibitem[]{} Kokubo, E., \& Ida, S.\ 
1996, Icarus, 123, 180 
\bibitem[]{} Kokubo, E., \& Ida, S.\ 
1998, Icarus, 131, 171 
\bibitem[]{} Laughlin, G., 
Steinacker, A., \& Adams, F.~C.\ 2004, \apj, 608, 489 
\bibitem[]{} Lubow, S.~H., \& 
Ogilvie, G.~I.\ 2001, \apj, 560, 997 
\bibitem[]{} Masset, F.\ 2000a, \aap, 141, 
165 
\bibitem[]{} Masset, F.~S.\ 2000b, Disks, 
Planetesimals, and Planets, 219, 75 
\bibitem[]{} Masset, F.~S., 
Morbidelli, A., Crida, A., \& Ferreira, J.\ 2006a, \apj, 642, 478 
\bibitem[]{} Masset, F.~S., D'Angelo, 
G., \& Kley, W.\ 2006b, \apj, 652, 730 
\bibitem[]{} McNeil, D., Duncan, M., 
\& Levison, H.~F.\ 2005, \aj, 130, 2884 
\bibitem[]{} Nelson, R.~P.\ 2005, \aap, 443, 
1067 
\bibitem[]{} Nelson, R.~P., \& 
Papaloizou, J.~C.~B.\ 2003, \mnras, 339, 993 
\bibitem[]{} Nelson, R.~P., \& 
Papaloizou, J.~C.~B.\ 2004, \mnras, 350, 849 
\bibitem[]{} Papaloizou, 
J.~C.~B., Nelson, R.~P., \& Snellgrove, M.~D.\ 2004, \mnras, 350, 829 
\bibitem[]{} Ogihara, M., Ida, S., 
\& Morbidelli, A.\ 2007, Icarus, 188, 522 
\bibitem[]{} Paardekooper, 
S.-J., \& Mellema, G.\ 2006, \aap, 459, L17 
\bibitem[]{} Papaloizou, 
J.~C.~B., \& Szuszkiewicz, E.\ 2005, \mnras, 363, 153 
\bibitem[Pi{\'e}tu et al.(2006)]{2006A&A...460L..43P} Pi{\'e}tu, V., 
Dutrey, A., Guilloteau, S., Chapillon, E., \& Pety, J.\ 2006, \aap, 460, 
L43 
\bibitem[]{} Pollack, J.~B., 
Hubickyj, O., Bodenheimer, P., Lissauer, J.~J., Podolak, M., \& Greenzweig, 
Y.\ 1996, Icarus, 124, 62 
\bibitem[]{} Shakura, N.~I. \& 
Sunyaev, R.~A.\ 1973. \aap, 24, 337.
\bibitem[]{} Steiner E.\ 1996, The Chemistry Maths Handbook. Oxford. 
\bibitem[]{} Stone J.M., Ostriker E.C. \& Gammie, C. F.\ 1998,
   {\apj, 508}, L99.
\bibitem[]{} Tanaka, H., Takeuchi, 
T., \& Ward, W.~R.\ 2002, \apj, 565, 1257 
\bibitem[]{} Tanaka, H., \& Ward, 
W.~R.\ 2004, \apj, 602, 388 
\bibitem[]{} Terquem, C., \& 
Papaloizou, J.~C.~B.\ 2007, \apj, 654, 1110 
\bibitem[]{} Turner, N.~J., Sano, T., 
\& Dziourkevitch, N.\ 2007, \apj, 659, 729 
\bibitem[]{} Ward, W.~R.\ 1997, Icarus, 126, 261 
\end{thebibliography}
\end{document}